\documentclass[twocolumn,aps,prb,showpacs,superscriptaddress,amsfonts,
amssymb,amsmath]{revtex4}
\usepackage{graphicx}
\begin{document}
\title{Universal spin-Hall conductance fluctuations in two dimensions}
\author{Wei Ren}
\affiliation{Department of Physics and Center of Theoretical and
Computational Physics, The University of Hong Kong, Hong Kong,
China}
\author{Zhenhua Qiao}
\affiliation{Department of Physics and Center of Theoretical and
Computational Physics, The University of Hong Kong, Hong Kong,
China}
\author{Jian Wang$^*$}
\affiliation{Department of Physics and Center of Theoretical and
Computational Physics, The University of Hong Kong, Hong Kong,
China}
\author{Qingfeng Sun}
\affiliation{Institute of Physics, Chinese Academy of Sciences,
Beijing, China}
\author{Hong Guo}
\affiliation{Department of Physics, McGill University, Montreal, PQ,
Canada}

\begin{abstract}
We report a theoretical investigation on spin-Hall conductance
fluctuation of disordered four terminal devices in the presence of
Rashba or/and Dresselhaus spin-orbital interactions in two
dimensions. As a function of disorder, the spin-Hall conductance
$G_{sH}$ shows ballistic, diffusive and insulating transport
regimes. For given spin-orbit interactions, a universal spin-Hall
conductance fluctuation (USCF) is found in the diffusive regime. The
value of the USCF depends on the spin-orbit coupling $t_{so}$, but
is independent of other system parameters. It is also independent of
whether Rashba or Dresselhaus or both spin-orbital interactions are
present. When $t_{so}$ is comparable to the hopping energy $t$, the
USCF is a universal number $\sim 0.18 e/4\pi$. The distribution of
$G_{sH}$ crosses over from a Gaussian distribution in the metallic
regime to a non-Gaussian distribution in the insulating regime as
the disorder strength is increased.
\end{abstract}
 \maketitle

The notion of dissipation-less spin-current\cite{murakami} has
attracted considerable interests recently. In its simplest form, a
spin-current is about the flow of spin-up electrons in one
direction, say $+x$, accompanied by the flow of equal number of
spin-down electrons in the opposite direction, $-x$. The total
charge current in the $x$-direction is therefore zero,
$I_e=e(I_{\uparrow}+I_{\downarrow})=0$; and the total spin-current
is finite: $I_s=\hbar/2(I_{\uparrow}-I_{\downarrow})\neq 0$. For a
pure semiconductor system with spin-orbital (SO) interactions, it
has been shown\cite{murakami} that an electric field in the
$z$-direction can induce the flow of a spin-current in the
$x$-direction perpendicular to the electric field: such a
spin-current is dissipation-less because the external electric field
does no work to the electrons flowing inside the spin-current. If
the semiconductor sample has a finite $x$-extent, the flow of
spin-current should cause a spin accumulation at the edges of the
sample, resulting to a situation that spin-up electrons accumulate
at one edge while spin-down electrons at the opposite edge. Hence a
spin-Hall effect\cite{hirsch,sinova} is produced where chemical
potentials for the two spin channels become different at the two
edges of the sample. This interesting phenomenon has been subjected
to extensive studies and there are several experiments reporting
spin accumulation which may have provided evidence of this
effect\cite{exp}. It has been shown that for a pure two dimensional
(2d) sample without any impurities, the Rashba SO interaction
generates a spin-Hall conductivity having universal value
\cite{sinova} of $e/8\pi$. It has also been shown that any presence
of weak disorder destroys this spin-Hall effect in the large sample
limit\cite{inoue,mishchenko}. On the other hand, numerical studies
have provided evidence that for mesoscopic samples, spin-Hall
conductance can survive weak disorder\cite{hank,sheng1,nikolic}.

One of the most striking quantum transport features in mesoscopic
regime is the universal charge conductance fluctuation
(UCF)\cite{web1,althu,lee85}: quantum interference gives rise to the
sample-to-sample fluctuation of charge conductance of order $e^2/h$,
independent of the details of the disorder, Fermi energy, and the
sample size as long as transport is in the coherent diffusive regime
characterized by the relation between relevant length scales,
$l<L<\xi$. Here $L$ is the linear sample size, $l$ the elastic mean
free path and $\xi$ the phase coherence length. If time-reversal
symmetry is broken, UCF is suppressed by a factor of two.

A very important and interesting issue therefore arises: what are
the properties of the fluctuations of {\it spin-Hall conductance} in
disordered samples? Is there a transport regime where spin-Hall
conductance fluctuation is {\it universal}? The answers to these
questions are non-trivial because the flow of dissipation-less
spin-current is qualitatively different from the flow of charge
current driven by an external electric field. It is the purpose of
this paper to report our investigations of these issues. For a
disordered four-terminal sample with a given Rashba SO interaction
strength $t_{so}$, and/or Dresselhaus interaction strength
$t_{so2}$, our results suggest that there is indeed a universal
spin-Hall conductance fluctuation (USCF) whose root mean square
amplitude is $g=0.18(e/4\pi)$, independent of other system details
(thus universal). The fluctuation is however a function of the SO
interaction strength and found to be well fitted by a functional
form of $\text{rms}(G_{sH})= g\tanh(|t_{so}-t_{so2}|/0.17)$.
Finally, the distribution of spin-Hall conductance obeys a Gaussian
distribution in the metallic regime and deviates from it in the
insulating regime.

\begin{figure}
\includegraphics[angle=-90,width=3.4in]{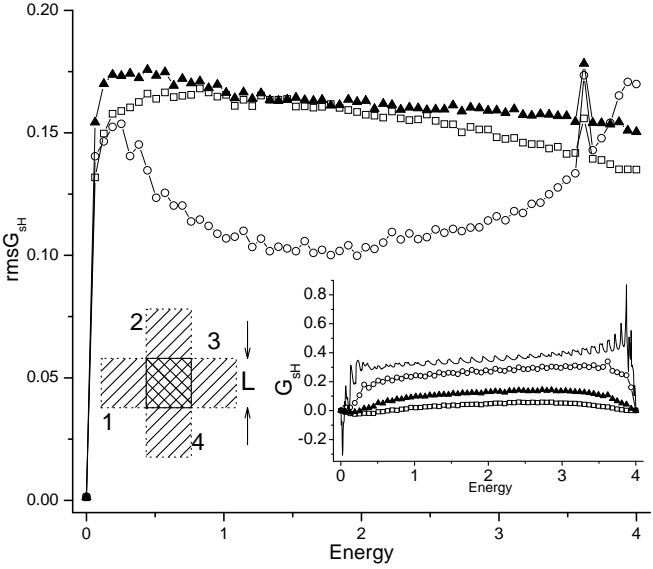}
\caption{ Spin-Hall conductance fluctuation versus energy for
disordered samples. Triangles, squares, and circles are for
$W=1,2,3$ respectively. Left inset: schematic plot of the
four-terminal mesoscopic sample where Rashba interaction exists in
the center scattering region and the leads 1, 3.  The width of the
square sample is $L$. A small voltage bias is across leads 1,3, and
spin-Hall conductance is measured through leads 2,4. Right inset:
the ensemble averaged spin-Hall conductance $G_{sH}$ versus electron
energy for SO interaction strength $t_{so}=0.3$. Solid line: pure
sample with $W=0$ and other symbols are the same as the main panel.
In all figures the spin conductance and its fluctuation are measured
in $e/4\pi$. } \label{fig1}
\end{figure}

To investigate USCF, we consider a four terminal device in two
dimensions schematically shown in the left inset of Fig.\ref{fig1}.
We will first discuss the results in the presence of only Rashba
interaction. In the presence of Rashba interaction ($\alpha_{so}
{\bf z}\cdot({\bf \sigma} \times {\bf k})$), the Hamiltonian of this
device is given by\cite{sheng1}:
\begin{eqnarray}
H=-t\sum_{<ij>\sigma}(c^\dagger_{i\sigma} c_{j\sigma}+h.c.)+\sum_{i\sigma}\epsilon_i c^\dagger_{i\sigma} c_{i\sigma} \nonumber \\
-t_{so}\sum_i [ (c^\dagger_{i,\uparrow} c_{i+{\hat x},\downarrow}
-c^\dagger_{i,\downarrow} c_{i+{\hat x},\uparrow}) \nonumber \\
-i( c^\dagger_{i,\uparrow} c_{i+{\hat y},\downarrow}
+c^\dagger_{i,\downarrow} c_{i+{\hat y},\uparrow})+h.c.] 
\end{eqnarray}
where $c^\dag_{i\sigma}$ is the creation operator for an electron
with spin $\sigma$ on site $i$, ${\hat x}$ and ${\hat y}$ are unit
vectors along x and y directions. Here $t=\hbar^2/2ma^2$ is the
hopping energy and $t_{so} = \alpha_{so}/2a$ is the effective
spin-orbit coupling. The on-site energy is given by
$\epsilon_{i}=4t$. In addition, static Anderson-type disorder is
added to $\epsilon_{i}$ with a uniform distribution in the
interval $[-W/2,W/2]$ where $W$ characterizes the strength of the
disorder. We consider the situation where Rashba interaction is present
everywhere except in leads 2 and 4 (see the left inset of Fig.\ref{fig1})
in order to measure the conserved spin-current\cite{sheng1}. We
apply external bias voltages at the four terminals as $(V_i,
i=1\cdots 4)=(v/2,0,-v/2,0)$: such a setup generates a
spin-current flowing from lead 2 to 4, {\it i.e.} a spin-Hall
effect measured from these two leads\cite{sheng1}

The spin Hall conductance $G_{sH}$ is calculated from the
Landauer-B\"{u}ttiker formula\cite{hank}
\begin{equation}
G_{sH}=(e/8\pi)[(T_{2\uparrow,1}-T_{2\downarrow,1})-(T_{2\uparrow,3}
-T_{2\downarrow,3})]
\label{GsH}
\end{equation}
where the transmission coefficient is given by $T_{2\sigma,1} ={\rm
Tr}(\Gamma_{2\sigma} G^r \Gamma_1 G^a)$, here $G^{r,a}$ are the
retarded and advanced Green's functions of central disordered region
of the device which we evaluate numerically. The quantities
$\Gamma_{i\sigma}$ are the line width functions describing coupling
of the leads to the scattering region, and are obtained by
calculating self-energies due to the semi-infinite leads using a
transfer matrices method\cite{lopez84}. The spin-Hall conductance
fluctuation is defined as $\text{rms}(G_{sH})\equiv
\sqrt{\left\langle G_{sH}^{2}\right\rangle -\left\langle
G_{sH}\right\rangle ^{2}}$, where $\left\langle
{\cdots}\right\rangle $ denotes averaging over an ensemble of
samples with different configurations of the same disorder strength
$W$. Note that in the presence of disorder, although one could use
another the definition $ {\bar
G}_{sH}=(e/4\pi)(T_{2\uparrow,1}-T_{2\downarrow,1})$ to calculate
and obtain the same {\it average} spin-Hall conductance as that of
$G_{sH}$, the spin-Hall fluctuation can only be obtained correctly
using the definition of Eq.(\ref{GsH}). We perform our calculations
on $L\times L$ square samples with four leads described above.
Sample sizes of $L=40$ up to 100 are examined\cite{foot5}. To fix
units, throughout this paper we measure the energy $E$, disorder
strength $W$, and spin-orbit coupling $t_{so}$ in terms of the
hopping energy $t$.

Fig.\ref{fig1} plots spin Hall conductance fluctuation vs Fermi
energy at a fixed $t_{so}=0.3$ and sample size $L=40$ for several
disorder strengths $W = 1,2,3$. We have also shown the averaged spin
Hall conductance in the right inset. Due to the electron/hole
symmetry, only data for energy range $[0,4]$ are shown. Over
$10~000$ disordered samples are averaged. For pure sample without
disorder, as the Fermi energy is increased the number of subbands
increase. As a result, the spin-Hall conductance $G_{sH}$ shows
small oscillations. When disorder is increased from zero, $G_{sH}$
decreases as expected and eventually, the small oscillation due to
the subbands vanishes. Most importantly, Fig.\ref{fig1} shows
substantial sample to sample fluctuations of $G_{sH}$, measured by
$\text{rms}(G_{sH})$, of the order $\delta \frac{e}{4\pi}$ where
$\delta$ is a number between $0.1$ and $0.2$. Such an amplitude of
fluctuation is comparable to the spin-Hall conductance itself.

\begin{figure}
\includegraphics[angle=-90,width=3.4in]{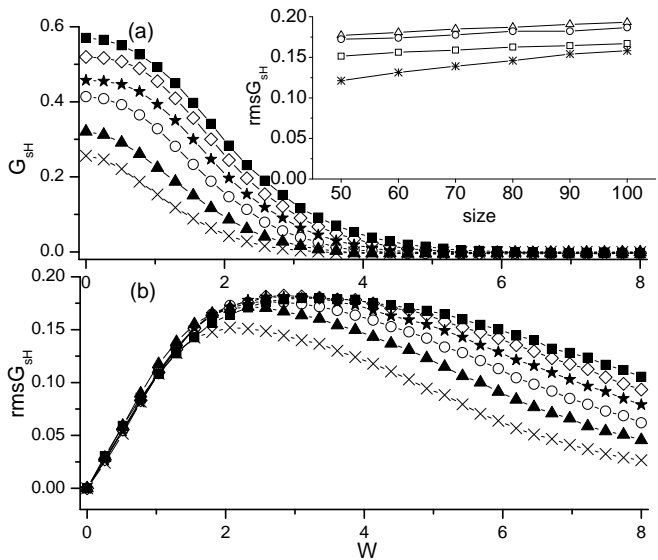}
\caption{ (a) Ensemble averaged $G_{sH}$ versus disorder strength
$W$ for $t_{so}=0.2$ (cross), 0.3 (filled triangle), 0.4 (open
circle), 0.5 (star), 0.6 (rhombus), 0.7 (filled square).
The average is over $20~000$ samples with $L=40$. Inset: size
dependence of spin-Hall conductance fluctuation with $t_{so}=0.3$.
Different symbols are for $W=1$ (stars), 2 (rectangles), 3 (circles)
and 4 (triangles). The ensemble average is over $20~000$ samples for
different size $L$. (b) The corresponding ensemble averaged
spin-Hall conductance fluctuation versus $W$, the symbols are for
the same $t_{so}$ values as in (a). } \label{fig2}
\end{figure}

In Fig.2a,b, we plot the $\langle G_{sH} \rangle$ and
$\text{rms}(G_{sH})$ as a function of disorder strength $W$ at fixed
Fermi energy $E=1$ for a number of different spin-orbit couplings
from $t_{so}=0.2$ up to $0.7$. Several observations are in order.
First, the spin-Hall conductance (Fig.\ref{fig2}a) decreases
smoothly as the disorder strength is increased: the transport
characteristics goes from quasi-ballistic at small $W$ to the
diffusive regime at larger $W$. In the diffusive regime, the
spin-Hall conductance decreases exponentially with the disorder
strength between $W=[1,5]$. Finally it goes to the insulating regime
for even larger $W$ where $G_{sH}$ vanishes. Second, the numerical
data show that the onsets of insulating regime $W_c$ are different
for different spin-orbit couplings $t_{so}$. The larger the
spin-orbit coupling, the larger $W_c$. This finding is consistent
with that of Ref.\onlinecite{sheng1} which suggested that the
localization length depends on $t_{so}$ and belongs to two-parameter
scaling. Third, from the spin-Hall conductance fluctuation shown in
Fig.\ref{fig2}b, we observe that for small disorder $W<1$, the
fluctuations for different $t_{so}$ have very similar values and the
curves collapse. For larger disorder, the fluctuations develop a
plateau structure so that $\text{rms}(G_{sH})$ becomes independent
of the disorder parameter $W$ for each given $t_{so}$. In this
sense, the fluctuation $\text{rms}(G_{sH})$ becomes ``universal''
and the spin-Hall transport enters the regime with universal
spin-Hall conductance fluctuations. Importantly, both the width of
the plateau and the value of USCF depend on $t_{so}$. The larger the
$t_{so}$, the wider the fluctuation plateau which characterizes the
diffusive regime.

\begin{figure}
\includegraphics[angle=-90,width=3.4in]{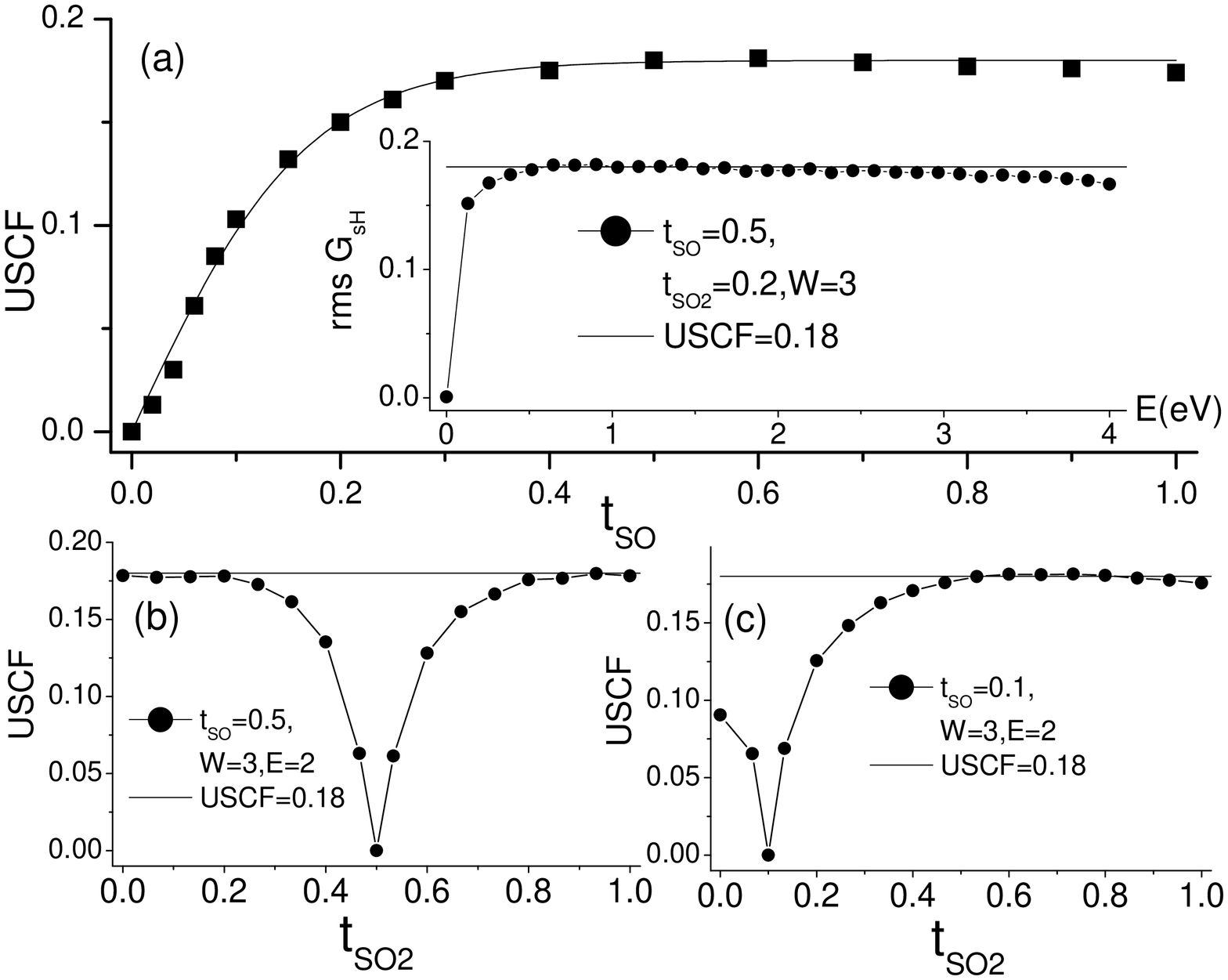}
\caption{ (a). USCF values versus Rashba spin-orbit coupling
$t_{so}$ at $E=1$. Inset: rms$G_{sH}$ versus Fermi energy for $W=3$
in the presence of both Rashba and Dresselhuss SO coupling,
$t_{so}=0.5$ and $t_{so2}=0.2$. (b)/(c). rms$G_{sH}$ versus
Desselhaus SO coupling $t_{so2}$ at $E=2$, $W=3$, and
$t_{so}=0.5$/$t_{so}=0.1$. In the cases of inset of (a), (b) and
(c), $40 000$ samples are collected.} \label{fig3}
\end{figure}

Now we examine the dependence of spin-Hall conductance fluctuation
$\text{rms}(G_{sH})$ on system size $L$ in the inset of
Fig.\ref{fig2}a for $t_{so}=0.3$. With weak disorder $W=1$ (stars)
the fluctuation increases with sample size indicating that spin-Hall
conductance is not yet in the USCF regime because transport is
quasi-ballistic. In the diffusive regime, $W=2,3,4$, the
fluctuations saturate at $\delta e/4\pi$ where $\delta \approx 0.2$.
The independence of system size by the fluctuation
$\text{rms}(G_{sH})$ provides a strong evidence of USCF. Namely, as
long as transport is in the diffusive regime, the fluctuation of the
spin-Hall conductance is dominated by quantum interference giving
rise to a universal amplitude. Since the value of USCF (which is
obtained from each curve in the lower panel of Fig.2) depends on the
spin-orbit coupling $t_{so}$ as seen from Fig.2, we have obtained a
collection of the USCF for different $t_{so}$ which is shown in
Fig.3a. Interestingly, the USCF can be well fitted (solid line) by a
function $\text{rms}(G_{sH}) = g\tanh(t_{so}/0.17)$ where $g=0.18
e/4\pi$. This can be understood as follows. When $t_{so}=0$, there
is no spin-Hall current and hence no fluctuations. There is a
crossover regime before the fluctuation saturates to the USCF
plateau.

\begin{figure}
\includegraphics[angle=-90,width=3.4in]{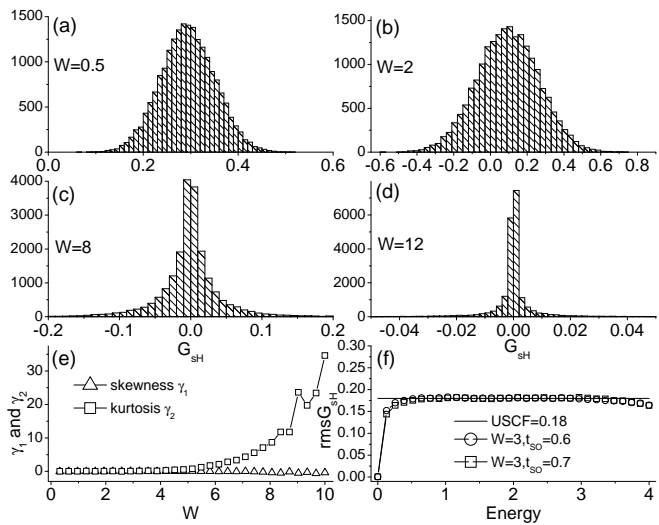}
\caption{ (a-d) The distribution of the spin-Hall conductance for
different disorder strengths at a fixed energy $E=1$ and
$t_{so}=0.3$. Data are collected for $20~000$ samples for each $W$
with $L=40$.  (e) The skewness $\gamma_1$ and kurtosis $\gamma_2$
versus disorder strength $W$ for the same ensemble. (f) rms$G_{sH}$
versus Fermi energy for $W=3$ and $t_{so}=0.6,0.7$.} \label{fig4}
\end{figure}

Fig.\ref{fig4}a-d plot the distribution function of spin Hall
conductance, $P(G_{sH})$, for several different disorder values $W$.
For each $W$, data is accumulated by calculating $20~000$
realizations of disorder. $P(G_{sH})$ appears to clearly obey a
Gaussian distribution in the metallic regime up to $W \approx 5$.
For larger $W$ between $[5,10]$, transport is in the insulating
regime, the symmetric distribution exhibits non-Gaussian behavior
(Fig.\ref{fig4}c). At even larger disorder $W=12$ shown in
Fig.\ref{fig4}d, the distribution becomes non-Gaussian and
asymmetric. The deviation from Gaussian distribution can be
characterized by the moments of spin-Hall conductance. We have
calculated the skewness $\gamma_1$ and kurtosis $\gamma_2$ whose
definitions are\cite{comment1},
$\gamma_{1}=\mu_{3}/\mu_{2}^{3/2}$
and
$\gamma_{2}=\mu_{4}/\mu_{2}^{2}-3$
where $\mu_{n}=\langle (x-\langle x \rangle)^n\rangle$ ($n=2,3,4$)
denote the central moments. The skewness describes the degree of
asymmetry of a distribution around its mean while the kurtosis
measures the relative peakedness of a distribution. The results are
plotted in Fig.\ref{fig4}f, showing that in the metallic regime
$W<5$, both skewness and kurtosis are essentially zero while they
become non-zero for larger $W$, consistent with the distributions.
Hence the skewness and kurtosis can be used to identify the
diffusive regime. Importantly, these quantities can be measured
experimentlly\cite{comment1}. Finally, we have checked that the
above features of the spin-Hall conductance fluctuation are generic
and valid for other values of $E$ and $t_{so}$. For instance, Fig.4f
shows the rms$G_{sH}$ versus Fermi energy when $W=3$ and
$t_{so}=0.6,0.7$. We see that between $E=1$ and $E=3$, rms$G_{sH}$
is around the universal value 0.18.

So far we have focused on spin-Hall conductance fluctuations with
the Rashba interaction. To further demonstrate the universal
behavior, we have also analyzed the case of  Dresselhaus
spin-orbital interaction by adding a term $\beta_{so} (\sigma_x k_x
- \sigma_y k_y)$ in Eq.(1). Using unitary transformations, it is
easy to prove that for the spin-Hall current along z-direction we
have $I_{sH}^z(\alpha_{so}=0,\beta_{so}) = I^z_{sH}
(\alpha_{so},\beta_{so}=0)$ and $I^z_{sH}(\alpha_{so}=\beta_{so}) =
0$. When there is no Rashba spin-orbital interaction
($\alpha_{so}=0$) and only Dresselhaus term exists, we have obtained
exactly the same USCF value and behavior as that of Rashba term
alone.  When both Rashba and Dresselhaus terms are present, it is
not obvious that the USCF persists. This is because these two terms
have different symmetry, the Rashba coupling arises from the
structure inversion asymmetry with SU(2) symmetry while the
Dresselhaus coupling arises from the bulk inversion asymmetry with
SU(1,1) symmetry\cite{loss}. From Fig.3b and 3c plot we see that
USCF for the latter situation (both $\alpha_{so}$ and
$\beta_{so}\neq 0)$, the results are similar to the case of pure
Rashba or pure Dresselhaus interaction.  Because
$I^z_{sH}(\alpha_{so}=\beta_{so}) = 0$, the USCF curves have a dip
to zero when $\alpha_{so}=\beta_{so}$. Defining $t_{so2} \equiv
\beta_{so}/2a$, Figs. 3b, 3c show that USCF is reached when
$|t_{so2}-t_{so}|\sim 0.4$. Finally, the inset of Fig.3a shows that
the USCF is independent of Fermi energy when both SO interactions
are present.

In summary, for spin-Hall effect generated by the Rashba and Dresselhaus
interactions in mesoscopic samples, our results strongly suggest the
existence of a universal spin-Hall conductance fluctuation due to
impurity scattering in the quantum coherent regime. In this regime,
the USCF is characterized by sample-to-sample fluctuations of
$G_{sH}$ for a given SO interaction (Rashba or/and Dresselhaus)
strength $t_{so}$, measured by
quantity $\text{rms}(G_{sH})$ with a {\it universal} amplitude
$g\equiv \delta \frac{e}{4\pi}$ where $\delta \approx 0.18$, which
is independent of system size, impurity scattering strength, and
Fermi energy. Importantly, this fluctuation amplitude is of the same
order as the spin-Hall conductance itself. Comparing with the
familiar UCF in charge conductance of disordered mesoscopic samples,
USCF originates from a similar physics of quantum interference
effect which leads to significant sample-to-sample fluctuations in
spin-Hall conductance. A main difference is that spin-Hall effect is
due to SO interactions in the sample, thereby the USCF also depends
on the SO parameter $t_{so}$ (or $t_{so2}$), and our numerical results
can be well fitted by a functional form of $\text{rms}(G_{sH})=
g\tanh(|t_{so}-t_{so2}|/0.17)$.

{\bf Acknowledgments} This work was financially supported by RGC
grant (HKU 7044/05P) from the government SAR of Hong Kong. Q.F.S is
supported by NSF-China under Grant No. 90303016 and 10474125 and H.G
is supported by NSERC of Canada, FQRNT of Qu\'{e}bec and Canadian
Institute of Advanced Research. Computer Centre of The University of
Hong Kong is gratefully acknowledged for the High-Performance
Computing assistance.

\noindent{$^{*)}$ Electronic address: jianwang@hkusub.hku.hk}


\begin{thebibliography}{99}
\bibitem{foot5}
Due to the extremely large computational demands for adequate
ensemble averaging, very careful coding must be done to exploit
the mathematical structure of the self-energy and Hamiltonian
matrices which allows the Green's function to be obtained by
inverting a tridiagonal matrix. For $L=40$, our code takes
approximately 10 minutes to calculate a curve of spin-Hall
conductance versus energy with 200 energy points on a single PC.
As tests, we have reproduced results in Fig.1-3 of
Ref.\onlinecite{sheng1}.
\end{thebibliography}
\end{document}